\documentclass[twocolumn,aps,pre,reprint,groupedaddress]{revtex4}
\usepackage{hyperref}
\usepackage{bigints}
\usepackage{amsmath, amssymb, graphicx, xcolor}
\usepackage{subfigure}

\begin{document}

\title{Detrimental role of fluctuations in the resource dependency networks}
\author{Saumitra Kulkarni}
\author{Snehal M. Shekatkar}
\email[]{snehal@inferred.in}
\affiliation{Department of Scientific Computing, Modeling and Simulation, Savitribai Phule Pune University, Pune 411007, India}

\begin{abstract}
    Individual components of many real-world complex networks produce and exchange resources among themselves. However, because the resource production in such networks is almost always stochastic, fluctuations in the production are unavoidable. In this paper, we study the effect of fluctuations on the resource dependencies in complex networks. To this end, we consider a modification of a threshold model of resource dependencies in networks that was recently proposed, where each vertex has a \emph{fitness} that depends on the total amount of resource it has produced, the amount it has procured from its neighbours, and the fitness threshold. We study how the ``network fitness'', defined as the average fitness of vertices in the network, is affected as the fluctuation size is varied. We show that the fluctuations worsen the network fitness even when average production on vertices is kept fixed. This is true independent of whether more than required amount is produced in the network or not. However, this effect saturates for large fluctuations, and hence very large fluctuations cannot worsen the network fitness beyond a limit.  We further show that the networks with a homogeneous degree distribution, such as the Erd{\H o}s-R{\'e}nyi network, are less affected by fluctuations and also produce lower wastage than the networks with a heterogeneous degree distribution like the Scale-Free network. Our work shows that fluctuations in the resource production should be avoided in resource dependency networks.
\end{abstract}


\maketitle

\section{Introduction}
Complex networks have emerged as a unifying framework to study the real-world complex systems that are made up of a large number of individual elements or units \cite{newman2018networks, albert2002statistical}. Components of most real-world complex system require adequate amount of one or more resources for proper functioning and for survival. Some common examples include water and food required by humans, electricity required by computers and routers, and raw material required by manufacturing firms. Usually the components need some threshold amount of a resource so that if the amount is less than the threshold of a component, its performance degrades or the component may die altogether. An important point of consideration is that a given component may not be capable of producing a required resource or may not produce it as much as it is required, and hence needs to procure some amount from other components of the system. For example, not every human being produces food material but can get it from other humans who do produce it. Similarly, most manufacturing firms do not produce their own electricity, and hence must buy it from entities like electricity boards. These links between components of a bigger system can be viewed as a complex network in which nodes produce various resources and share the surplus amounts to others which need those \cite{Ingale_Shekatkar2020, agrawal2022effect}. Trade networks, in which resources are exchanged for money or other resources is a classic example of such network \cite{Garlaschelli_WTW, Vece_gravity, garlaschelliFitnessDependentTopologicalProperties2004, squartiniRandomizingWorldTrade2011a, squartiniRandomizingWorldTrade2011, diveceGravityModelsNetworks2022}.

In the manufacturing industry, supply chains are of utmost importance. Fundamentally, supply chains are networks of firms or manufacturing entities in which raw materials are procured and converted into ancillary or final products. There have been many attempts of modelling the dynamics of supply-chains \cite{swaminathan1998modeling, de2000supply}.  The goal of such modelling efforts is to try to maximize their individual profit and livelihood \citep{choiSupplyNetworksComplex2001, ivanovRippleEffectSupply2014, swierczekImpactSupplyChain2014, shaoDataanalyticsApproachIdentifying2018, diemQuantifyingFirmlevelEconomic2022b}.

A related network type in which some resource flows from one vertex to another is distribution networks. Some examples of distribution networks are power-grids \cite{carrerasModelingBlackoutDynamics2001, brummittSuppressingCascadesLoad2012}, networks of gas pipelines \cite{carvalhoRobustnessTransEuropeanGas2009}, river networks \cite{maritanScalingLawsRiver1996, doddsGeometryRiverNetworks2000, doddsGeometryRiverNetworks2000a, rinaldoTreesNetworksHydrology2006}, cardiovascular and respiratory networks \cite{westGeneralModelOrigin1997, banavarSizeFormEfficient1999}, and plant vascular and root systems \citep{enquistAllometricScalingPlant1998, damuthCommonRulesAnimals1998, westGeneralModelStructure1999}. 

All these networks mentioned above fall under a more general category of networks called the \emph{resource dependency networks} in which the vertices depend on each other for various resources. In this work, we aim to advance the work related to such networks that have recently been studied from a perspective of complex networks \cite{Ingale_Shekatkar2020, agrawal2022effect}. To this end, we propose a modified version of the model introduced in \cite{Ingale_Shekatkar2020} which we call the \emph{Suplus Distribution Model}. We particularly want to find out how the fluctuations in the production can affect the quality of the state of the network as quantified by \emph{Network Fitness} in our work. A major difference between these previous studies and the present one is that when a vertex does not have a threshold amount, instead of dying, only its fitness reduces. This is true in most real-world scenarios. In this study, we also introduce the notion of \emph{Wastage}, and systematically study effects of network topology and stochasticity of the resource production on it. 

The rest of the paper is organized as follows. In Sec.\ref{surplus_model} we present a variation of the model of resource dependency in \cite{Ingale_Shekatkar2020}. In this section, apart from the model description, we also describe the way we are using different probability distributions to generate resource at each vertex as well as the process of generating substrate networks to carry out simulations. In Sec.\ref{simulation_results}, we describe the simulation results obtained by varying the fluctuation size in the resource production. After that in Sec.\ref{wastage} we describe how the fluctuations affect the amount of resource wastage in a network, and we conclude in Sec.\ref{conclusion}.

\section{Surplus distribution model \label{surplus_model}}
\begin{figure*}[ht]
    \includegraphics[width=0.9\textwidth]{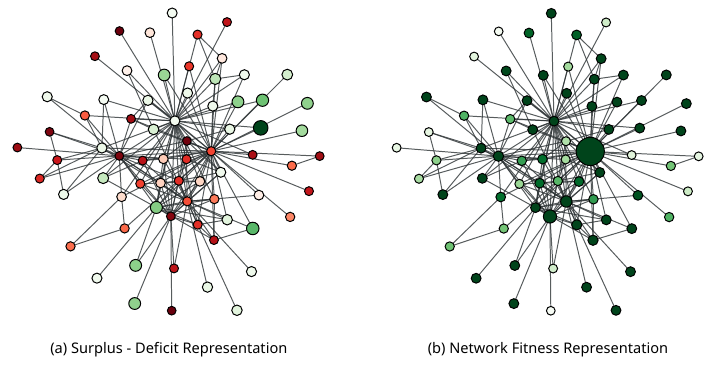}
    \caption{\label{graphic} Graphical representation of the \emph{Surplus Distribution Model}. (a) Each vertex $i$ requires resource amount $R_i$ and it produces amount $X_i$. The vertices for which $X \geq R$ are shown in green (higher surplus vertices darker) while the others are shown in red (higher deficit vertices darker). Every vertex with $X > R$ now shares its surplus $X-R$ with its neighbours proportional to their deficit. The total amount $X^{\text{tot}}$ on each vertex is thus the sum of the amount produced plus the amount received from the neighbours. (b) Vertices color coded according to their fitness which is $(X^{\text{tot}}/R)$ if $X^{\text{tot}} < R$ otherwise is $1$. In this picture, the sizes of vertices approximately indicate the amount present on them. }
\end{figure*}
The model that we use here is a variation of the model of resource dependencies proposed in \cite{Ingale_Shekatkar2020}. Consider a network with $n$ vertices and $m$ edges. Our model is agnostic to whether the network is directed or not, however all the simulations in this paper are done for undirected networks. Each vertex $i$ in the network stochastically produces amount $X_i(t)$ at each discrete time $t=0, 1, 2, \dots$. This amount $X_i$ is a random variable with a given probability distribution $p(x,\beta_i)$ where $\beta_i$ is the collection of parameters of the distribution. Also, each vertex $i$ requires a threshold amount $R_i$ of resource at each time step. 

Consider a vertex $j$ that may or may not be a neighbour of vertex $i$. If $X_j(t) > R_j$, vertex $j$ has \emph{surplus} $S_j(t)=X_j(t)-R_j$, which it distributes among all its neighbours. If $X_j(t) < R_j$, there is no surplus ($S_j(t)=0$), and consequently no sharing happens. The fraction of the surplus that is received by vertex $i$ depends linearly on its deficit. The deficit of vertex $i$ at time $t$, is $D_i(t)=R_i-X_i(t)$ if $X_i(t) < R_i$, otherwise it is zero. Thus the total amount on any vertex $i$ at time $t$ can be written as:

\begin{equation}
    \label{eq:X_tot}
    \begin{aligned}
        X_i^{\text{tot}}(t) = \begin{cases} X_i(t) + \sum\limits_{j=1}^n A_{ij}S_{j}(t)\frac{D_i(t)}{\sum\limits_{l=1}^{n}A_{lj}D_l}\quad \text{if}\ X_i(t) < R_i\\ 
\\
R_i \quad\quad \text{Otherwise}\end{cases}
    \end{aligned}
\end{equation}

Here $A_{ij}$ denotes the $(i,j)^{\text{th}}$ element of the adjacency matrix of the network.

An important aspect of the model is that the resource is assumed to have a lifetime of only $1$ unit of time. This assumption is valid for many perishable resources in the real-world which include agricultural goods like vegetables, fruits etc. and dairy products like milk. In line with this, we assume that if $X^{\text{tot}}_i(t) < R_i$, the vertex $i$ completely consumes it at time $t$. But if $X^{\text{tot}}_i(t) \geq R_i$, the vertex $i$ only consumes the amount $R_i$, and the remaining amount $X^{\text{tot}}_i(t)-R_i$ is discarded at time $t$, and none is left for time $t+1$. We define the fitness of a vertex $i$ at time $t$ as: 
\begin{equation}
f_i(t) = \min(X_i^{\text{tot}}(t)/R_i, 1)
\end{equation}

Fig.~\ref{graphic} shows a graphical representation of this model.\\

Let us denote the probability distribution of $X_i^{\text{tot}}$ by $Q_i(x)$. Then the expected value of the fitness of vertex $i$ is given by:
\begin{equation}
\label{eq:Qx}
\langle f_i\rangle = \int_{0}^{R_i}\frac{x}{R_i}Q_i(x)dx + \int_{R_i}^{\infty} Q_i(x)dx
\end{equation}
In the second term in the equation above, we don't multiply $Q_i(x)$ by $x/R_i$ because for any value of $x > R_i$, fitness is assumed to be $1$, which the maximum fitness a vertex can have. This means that for vertex $i$ to be maximally fit, it only needs to somehow have resource amount $R_i$  with it, and all amount greater than that is useless. Note that, in general, the distribution $Q_i(x)$ for the vertex $i$, apart from depending upon $p(x;\beta_i)$, would also depend on the topological characteristics of the vertex like its degree, centrality, clustering coefficient etc. Thus, even when all the parameters $\beta_i$ of the producing distribution are same, the distribution $Q(x)$ and hence the expected fitness, are in general different for different vertices. Moreover, since the network state at time $t+1$ is completely independent of the state at time $t$, the distribution $Q_i(x)$ and $f_i$ are time-independent. This also means that the parameter $t$ is redundant in our model. But keeping this parameter makes the model flexible so that it might be possible to extend it to Markovian dynamics or even to non-Markovian dynamics. 

A quantity of interest to us is the average expected fitness of vertices in the network, which in the limit of an infinite sized network is given by:
\begin{equation}
\label{eq:fitness}
F = \lim\limits_{n\to\infty} \frac{1}{n}\sum\limits_{i=1}^n\langle f_i\rangle
\end{equation}

We call $F$ the \emph{Network fitness}, and in the rest of the paper, we investigate how it is affected by the network topology and the fluctuations in the resource production on vertices. 

Since the quantity $F$ in Eq(\ref{eq:fitness}) is defined for an infinite sized network, we must find a way to estimate its value computationally using an ensemble of only finite sized networks. To do that, let us consider the following estimator for $F$:
\begin{equation}
    \widehat{F}(t) = \frac{1}{n}\sum\limits_{i=1}^n f_i(t)
\end{equation}
Because the networks we use to simulate the model contain only a finite number $n$ of vertices, this quantity would fluctuate with time. However, as $n$ increases, the fluctuations in $\widehat{F}$ would reduce in size. Because $Q_i(x)$ is time-independent, we can modify the estimator given above to include the time average as follows:
\begin{equation}
    \label{eq:F_estimator}
    \widehat{F} = \frac{1}{nT}\sum\limits_{t=0}^{T}\sum\limits_{i=1}^n f_i(t)
\end{equation}
where $T$ is the total number of time steps for which we iterate our model. Since now we have average over more replications, this estimator estimates $F$ more precisely. A central question that we investigate in this paper is whether for a given network topology, we can maximize $F$ by changing the size of fluctuations in the produced amount. 

\subsection{Choice of probability distributions \label{sec:choice_prob_dists}}
\begin{figure}[ht]
\includegraphics[width=0.9\columnwidth]{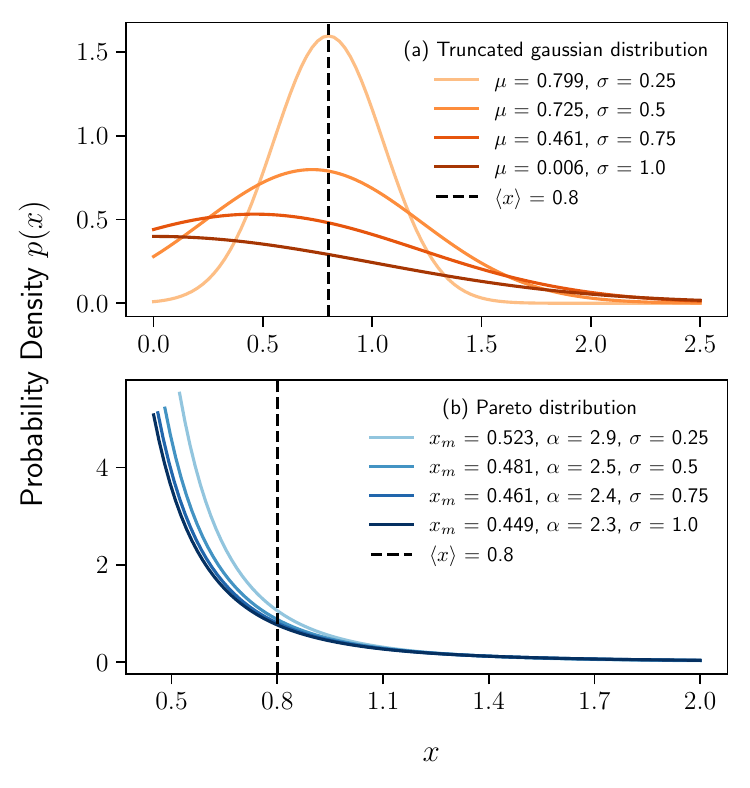}
    \caption{\label{fig:prob_distro} Probability density curves for the truncated Gaussian and the Pareto distributions for different sizes of fluctuation $\sigma$ with the mean of the distribution $\langle x\rangle$ fixed at $0.8$. (a) For the truncated Gaussian, the parameter $\mu$ decreases as $\sigma$ increases. (b) For the Pareto distribution, both $x_m$ and $\alpha$ decrease as $\sigma$ increases. Vertical dashed line marks the mean $\langle x\rangle$ of the distribution in both the subplots.}
\end{figure}
In this paper, we are interested in studying the effect of fluctuations, and for this, we must choose those distributions in which the fluctuation size can be varied keeping the mean fixed. Henceforth we will use the term \emph{resource generator} to refer to a probability distribution that is used to stochastically generate resource at each vertex. In the resource dependency model proposed in \cite{Ingale_Shekatkar2020}, the resource generator was the \emph{exponential} distribution $p(x) = \text{Exp}(x,\beta)$. However, for the exponential distribution, the mean and the standard deviation are equal, and hence it is not possible to vary the fluctuation size independent of mean. Thus, it was not possible there to separate the effects of the average production and the fluctuation size. For this reason, here we use Gaussian and Pareto distributions as resource generators. These two serve as prototypes of the classes of peaked and heavy-tailed distributions respectively. 

The usual Gaussian distribution has the whole real line as its support. However, the resource amount cannot be negative. Hence we use the \emph{truncated} Gaussian distribution which is obtained by removing the negative tail of the usual Gaussian distribution. The probability density of this distribution is given by:
\begin{equation}
    p(x; \mu, \sigma) = \frac{1}{\psi(\mu, \sigma)}\exp\left(-\frac{(x-\mu)^2}{2\sigma^2}\right)
\end{equation}

where, 
\begin{equation}
\psi(\mu, \sigma) = \int\limits_{x=0}^{\infty} \exp\left(-\frac{(x-\mu)^2}{2\sigma^2}\right)dx
\end{equation}

In the case of usual Gaussian distribution $\mathcal{N}(\mu,\sigma^2)$, the mean of the distribution $\langle x\rangle$ is equal to $\mu$, and hence is independent of $\sigma$. However, for the truncated Gaussian distribution, the mean is a function of both $\mu$ and $\sigma$ because truncation makes the distribution asymmetrical. In fact, the mean is given by:
\begin{equation}
\label{eq:truncated_gaussian_mean}
\langle x\rangle = \frac{1}{\psi(\mu, \sigma)}\int\limits_{x=0}^{\infty} x\exp\left(-\frac{(x-\mu)^2}{2\sigma^2}\right)dx
\end{equation}

To vary the fluctuation size, we can vary $\sigma$. However, if we do that keeping $\mu$ at a fixed value, the mean $\langle x\rangle$ also changes: increasing $\sigma$ leads to a flatter distribution resulting in the higher value of $\langle x\rangle$. For this reason, while increasing $\sigma$, we simultaneously decrease $\mu$ so that the mean remains fixed. This is an important consideration because we specifically want to study how changing the fluctuation size affects the network fitness for the same value of the average production. Since the integral in Eq(\ref{eq:truncated_gaussian_mean}) cannot be computed analytically, we numerically integrate it for a given value of $\sigma$ by systematically varying $\mu$ so as to find the value for which the R.H.S. of Eq(\ref{eq:truncated_gaussian_mean}) is equal to a given value of $\langle x\rangle$ within a numerical error. 

For the Pareto distribution, we have to use a similar strategy. The Pareto distribution is a continuous probability distribution with support $[x_m, \infty)$ where the parameter $x_m>0$. The density of the Pareto distribution is:
\begin{equation}
p(x; x_m, \alpha) = \frac{\alpha x_m^{\alpha}}{x^{\alpha+1}}
\end{equation}

where $\alpha$ is the shape parameter which decides how steep the distribution is. For this distribution, the mean exists only for $\alpha>1$, and is the function of both $x_m$ and $\alpha$:
\begin{equation}
\label{eq:pareto_mean}
\langle x\rangle = \frac{\alpha x_m}{\alpha-1}
\end{equation}

Also, for the Pareto distribution the standard deviation exists only for $\alpha > 2$, and is given by:
\begin{equation}
    \label{eq:pareto_sigma}
\sigma = \sqrt{\frac{\alpha x_m^2}{(\alpha-1)^2(\alpha-2)}}
\end{equation}

Eliminating $x_m$ from Eq(\ref{eq:pareto_mean}) and Eq(\ref{eq:pareto_sigma}), we get:
\begin{equation}
    \label{eq:pareto_alpha}
    \alpha = 1 + \sqrt{1+\left(\frac{\langle x\rangle}{\sigma}\right)^2}
\end{equation}

Thus, to vary the fluctuations, we vary $\sigma$ in the range $(0, 1)$ for a given $\langle x\rangle$. For each value of $\sigma$, we get a unique value of $\alpha$ from this equation. Then the Eq(\ref{eq:pareto_mean}) gives us $x_m$ corresponding to these $\sigma$ and $\langle x\rangle$ which can then be used to simulate the model. It is worth noting that as $\sigma$ increases, both $x_m$ and $\alpha$ decrease. This fact will be useful to us in the discussion ahead.

Thus for both truncated-Gaussian and Pareto generators, we use $\langle x\rangle$ as a parameter of the model which can be fixed to a suitable value, and then we vary the fluctuation size by changing $\sigma$ to see how the network fitness is affected. We present the simulation results in Sec.\ref{simulation_results}.

\subsection{Generation of substrate networks}
Apart from the parameters of the producing distribution $p(x;\beta_i)$, the Network Fitness $F$ is also affected by the network topology. Before we describe how we generate the substrate networks, we would like to clarify the meaning of a network in the present context. In the real-world, resources often flow between the vertices via physical edges like roads or pipelines. Thus, the distribution of the resource produced on a given vertex may reach not to just its first neighbours but also to its distant neighbours. However, in our context, any two vertices that share resources with each other are said to be connected by an edge. Thus, if in the original network the resource flows between the vertices $V_1$ and $V_3$ via the vertex $V_2$, but $V_1$ and $V_3$ are not connected, in our network, $V_1$ and $V_3$ would be connected since there is an exchange of resource between them. Thus, in our network, by construction the resource reaches only the first neighbour, and this is not a shortcoming of the model. 

In this paper, we restrict ourselves to studying only the degree distribution of the network. In particular, we want to see how the fluctuations in the production of resources affect the network fitness for two typical classes of degree distributions: peaked and heavy-tailed. As archetypal examples of these two classes, we choose Poisson and Power-law degree distributions. The configuration model of networks is a random graph model with a given degree sequence. However, in the limit of large network size ($n\to\infty$), this model can be thought of as a random graph model with a given degree distribution if a degree sequence is drawn from that distribution \cite{Newman2001}. As mentioned above, in this paper we are only interested in studying the effect of degree-distribution on the network fitness, and hence we use the configuration model with Poisson and Power-law degree distributions to generate the substrate networks. The corresponding networks are also known as the Erd{\H o}s-R{\'e}nyi network and the Scale-free network respectively. Henceforth, in this paper we refer to these two networks as ER and SF networks respectively. The only problem with the use of the configuration model is that it also allows multi-edges and self-loops. In our simulations, we remove all self-loops and we collapse all the multi-edges to single weighted edges with weights given by corresponding multiplicities. We then make the fraction of surplus shared along each edge proportional to its weight. 

It is important to make sure while comparing results for ER and SF networks that they have the same average degree. If $k_{\text{min}}$ denotes the minimum degree value in a SF network, its normalized power-law degree distribution has the form:
\begin{equation}
    p_k = \frac{k^{-\gamma}}{\zeta(\gamma, k_{\text{min}})}\quad\quad \text{for}\quad k \geq k_{\text{min}}
\end{equation}

\noindent while $p_k = 0$ for $k < k_{\text{min}}$. Here $\gamma$ is the scaling index of the power law and $\zeta(\gamma, k_{\text{min}})$ is the Hurwitz zeta function. If we draw degree values from this distribution, the value of the average degree is given by: 
\begin{equation}
    \langle k\rangle_{SF} = \sum\limits_{k=k_{\text{min}}}^{\infty} kp_k = \sum\limits_{k=k_{\text{min}}}^{\infty} \frac{k^{1-\gamma}}{\zeta(\gamma, k_{\text{min}})}
\end{equation}

In this paper, we choose $k_{\text{min}} = 2$ and $\gamma = 2.2$ which gives $\langle k\rangle_{SF} \approx 9.36$. To make the right comparison with the SF network, we sample the degree values for the ER network from the Poisson distribution: 
\begin{equation}
    p_k = e^{-\langle k\rangle_{SF}}\frac{\langle k\rangle_{SF}^k}{k!}
\end{equation}
This way, both types of networks theoretically have the same average degree. 

\section{Effect of fluctuations}
\label{simulation_results}
In this paper, we assume that the average production $\langle x\rangle$ is same for all the vertices. We also assume that the thresholds $R_i$ have the same value $R=1$ for all the vertices. Unless mentioned otherwise, all the numerical simulation results presented in this paper are obtained by averaging over $100$ realizations of networks each of size $n=10^4$. Also, for each realization, the model is simulated for $T=1000$ time steps. All the codes used in this paper are freely available as a part of the Python library \emph{dependency-networks} \cite{shekatkar2020dependency-networks}. 

\subsection{Effect of generators \label{sec:generators}}
\begin{figure}[ht]
    \includegraphics[width=0.95\columnwidth]{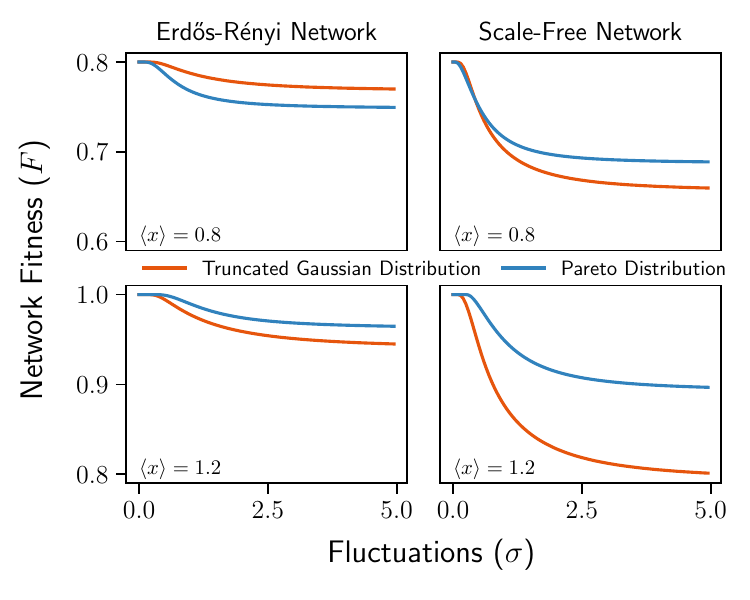}
    \caption{\label{fig:fitness} Variation of network fitness with fluctuation size for different combinations of network topologies, resource generators and average productions. \emph{Left} and \emph{Right} columns correspond to ER and SF topologies, while the \emph{Top} and \emph{Bottom} rows correspond to $\langle x\rangle < R$ and $\langle x\rangle > R$ respectively. In all the cases, increasing fluctuations can be seen to reduce network fitness even though average resource production is kept fixed. See text for a detailed discussion. }
\end{figure}
As we describe in this section, the type of generator (Gaussian vs Pareto) has a significant impact on the network fitness. To systematically study it, we consider the cases $\langle x\rangle < R$ and $\langle x\rangle > R$ separately. The results that we present below are obtained by varying fluctuation size $\sigma$ for various combinations of the generator and network topology when $\langle x\rangle < R$ and $\langle x\rangle > R$, which are shown in the top and bottom panels of Fig.~\ref{fig:fitness}. 

As the figure shows, independent of the generator and the network topology used, increasing the fluctuation size $\sigma$ worsens the network fitness $F$. This can be understood by the following argument. As $\sigma$ increases starting from zero, resource amounts greater than as well as less than $\langle x\rangle$ are generated. But since all resource amounts greater than $R$ lead to the same fitness $1$, amounts less than $\langle x\rangle$ dominate the average over vertices leading to the overall decrease in the network fitness. However, as the plots also show, increasing $\sigma$ does not decrease the network fitness beyond a certain value and the fitness saturates as $\sigma\to\infty$.

The same figure shows a peculiar fact that only for the combination of $\langle x\rangle < R$ and ER topology, asymptotically the Gaussian generator outperforms the Pareto generator while for all other combinations, the Pareto generator can be seen to perform better. 

\subsection{Effect of network topology \label{topo_effect}}
\begin{figure}[ht]
    \includegraphics[width=\columnwidth]{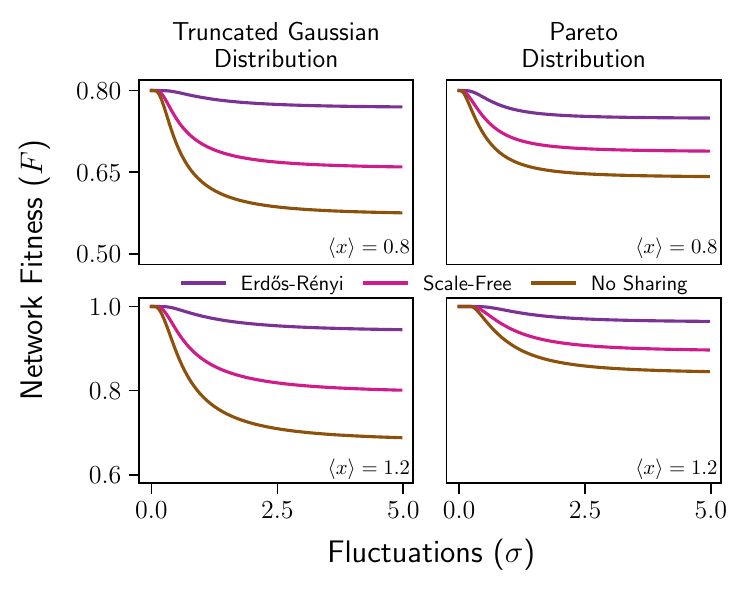}
    \caption{\label{fig:net_topology_effect} Comparison of network fitness curves of ER and SF topologies for various combinations of average productions and resource generators. The brown curve in each subplot shows the lower bound $F_L$ obtained analytically by considering a network in which vertices are not allowed to share their surplus amounts. \emph{Left} and \emph{Right} columns correspond to the two different generators, while the \emph{Top} and \emph{Bottom} rows correspond to $\langle x\rangle < R$ and $\langle x\rangle > R$ respectively. }
\end{figure}

In this section, we discuss the effect of network topology on the fitness. In Fig.~\ref{fig:net_topology_effect}, we compare $F$ as $\sigma$ is varied, for both ER and SF networks where different panels correspond to different combinations of $\langle x\rangle$ and generator. For a given combination, changing the network topology changes the quantitative behavior of $F$, but not the qualitative behavior. We also see that irrespective of the combination used, the ER network performs better than the SF network. We note here that a similar conclusion has been reached in the earlier works too \cite{Ingale_Shekatkar2020, agrawal2022effect}. To separately understand the effect of network topology on the fitness for the two generators, let us first consider a case in which the vertices are not allowed to share their surplus with other vertices. The fitness $F_{L}$ corresponding to this \emph{No-sharing} case is the lower bound for fitness for any network where sharing is allowed. It is easy to see that $F_{L}$ is same as $\langle f_i\rangle$ when $Q_i(x)$ in Eq(\ref{eq:Qx}) is replaced by $p(x;\beta_i)$:
\begin{equation}
F_L = \int_{0}^{R}\frac{x}{R}p(x;\beta)dx + \int_{R}^{\infty} p(x;\beta)dx
\end{equation}

The value of $F_{L}$ for the Gaussian generator is given by:
\begin{equation}
\label{eq:FNS_gauss}
    \begin{aligned}
    F_{L}^{gauss} &= \frac{1}{\psi(\mu, \sigma)}\int\limits_{x=0}^{R} \frac{x}{R}\exp\left(-\frac{(x-\mu)^2}{2\sigma^2}\right)dx \\
        &+ \frac{1}{\psi(\mu, \sigma)}\int\limits_{x=R}^{\infty} \exp\left(-\frac{(x-\mu)^2}{2\sigma^2}\right)dx
    \end{aligned}
\end{equation}
To compute $F_L$ as a function of $\sigma$, as described in Sec.\ref{sec:choice_prob_dists}, we compute the value of the parameter $\mu$ for a given $\sigma$ such that the average production $\langle x\rangle$ remains at a fixed value. We then compute the integrals in this equation numerically using this obtained value of $\mu$ for a given $\sigma$. 

Also, we can compute $F_L$ for the Pareto generator completely analytically as we describe now. Rearranging Eq(\ref{eq:pareto_mean}) for $x_m$, we get:
\begin{equation}
\label{eq:pareto_xm}
x_m = \left(1-\frac{1}{\alpha}\right)\langle x\rangle
\end{equation}

\begin{figure}[ht]
    \includegraphics[width=\columnwidth]{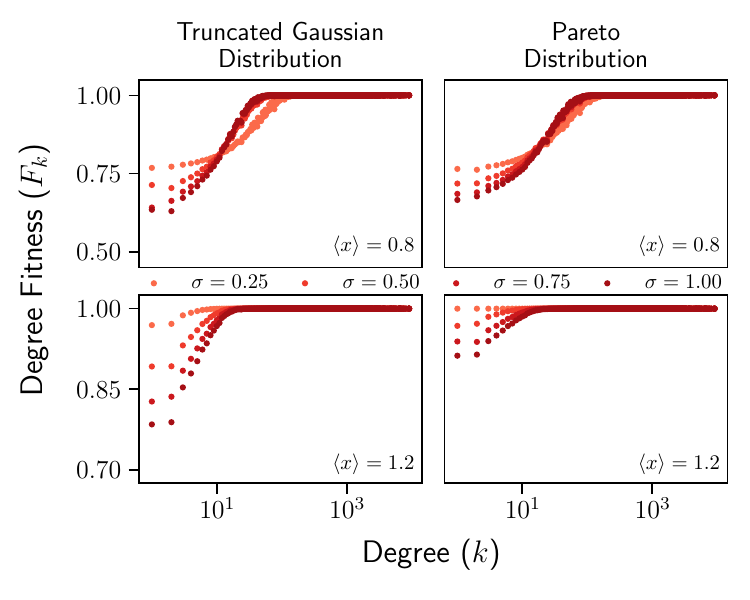}
    \caption{\label{deg_corr} Scatterplots of \emph{degree fitness} (average fitness of vertices of a given degree) and degree for three different fluctuation sizes for the SF network. \emph{Left} and \emph{Right} columns correspond to the two different generators, while the \emph{Top} and \emph{Bottom} rows correspond to $\langle x\rangle < R$ and $\langle x\rangle > R$ respectively. Although the quantitative behavior of the plots is different for different combinations of generators and average productions, it can be seen that the fitness tends to increase with degree. }
\end{figure}
When $\langle x\rangle < R$, this equation implies $x_m < R$ for all values of $\alpha$. Hence in this case:
\begin{equation}
    \label{eq:FL_pareto}
    \begin{aligned}
    F_L^{pareto} &= \int\limits_{x_m}^{R} \frac{\alpha x_m^{\alpha}}{x^{\alpha}}dx + \int\limits_{R}^{\infty} \frac{\alpha x_m^{\alpha}}{x^{\alpha+1}}dx \\ 
    &= \frac{\alpha x_m^{\alpha}}{1-\alpha}(R^{1-\alpha}-x_m^{1-\alpha}) + x_m^{\alpha}R^{-\alpha}
    \end{aligned}
\end{equation}
If needed, $F_L^{pareto}$ can be rewritten in terms of $\langle x\rangle$ and $\sigma$ using Eq(\ref{eq:pareto_xm}) and Eq(\ref{eq:pareto_alpha}). Then, it can be shown in a straightforward manner that 
\begin{equation}
    \lim\limits_{\sigma\to\infty} F_L^{pareto} = \langle x\rangle + \frac{\langle x\rangle^2}{R}\left(\frac{1}{4R}-\frac{1}{2}\right)
\end{equation}
From Eq(\ref{eq:pareto_xm}), when $\langle x\rangle > R$, whether $x_m < R$ or $x_m > R$ depends on the value of $\alpha$. Therefore there exists a transition value $\alpha_c$ such that for $\alpha < \alpha_c \implies x_m < R$ while $\alpha > \alpha_c \implies x_m > R$. For the transition value $\alpha_c$, $x_m = R$ and Eq(\ref{eq:pareto_xm}) gives:
\begin{equation}
R = \left(1-\frac{1}{\alpha_c}\right)\langle x\rangle \implies \alpha_c = \frac{\langle x\rangle}{\langle x\rangle-R}
\end{equation}

Using this $\alpha_c$ in Eq(\ref{eq:pareto_alpha}), we get the corresponding transition value $\sigma_c$:
\begin{equation}
    \alpha_c = \frac{\langle x\rangle}{\langle x\rangle-R} = 1 + \sqrt{1+\left(\frac{\langle x\rangle}{\sigma_c}\right)^2}
\end{equation}

which gives:
\begin{equation}
    \sigma_c = \sqrt{\frac{\langle x\rangle(\langle x\rangle-R)^2}{2R-\langle x\rangle}}
\end{equation}

For $\sigma < \sigma_c \implies x_m > R$, and hence the lower bound $F^{pareto}_L$ is given by:
\begin{equation}
    F^{pareto}_L = \int\limits_{x_m}^{\infty} p(x; x_m,\alpha)dx = 1
\end{equation}
Whereas, for $\sigma > \sigma_c$, $F_L^{pareto}$ is given by Eq(\ref{eq:FL_pareto}). Thus, we see that unlike the case $\langle x\rangle < R$, here we have two distinct regions in which qualitatively different behaviors are seen. In the region $\sigma < \sigma_c$, the network is in a completely fit state ($F=1$) whereas as soon as $\sigma$ crosses $\sigma_c$, it goes into a partially fit state ($F < 1$). 

The brown curve in each subplot in Fig.~\ref{fig:net_topology_effect} shows the variation of $F_L$ with $\sigma$. The plots in the figure also show that the actual shape of $F$ is not qualitatively very different from $F_L$. This means, the only effect of sharing is to scale the curves up. Furthermore, it is also clear that sharing smoothens the transition at $\sigma_c$ that is observed for the Pareto generator.

The fitness of a network is the average of fitness values of all the vertices in the network. However, the vertices are not equivalent in terms of fitness because they are not equivalent in terms of topology. In fact, it is clear from Eq(\ref{eq:X_tot}) that a vertex receives amount approximately proportional to its degree. Because of this, the high-degree vertices tend to be fit more often than the low-degree vertices. To verify this, we define the quantity \emph{Degree fitness} $F_k$ which is just the average fitness of a vertex with degree $k$. In Fig.~\ref{deg_corr}, we show scatterplots of $F_k$ and $k$ for various combinations of $\langle x\rangle$, $\sigma$, and the generator, for the SF network. As is evident from these plots, independent of the choice of these parameters, fitness is a rapidly increasing function of the degree, and for relatively small values of degree, the degree fitness reaches its maximum value $1$.
\section{Wastage \label{wastage}}
\begin{figure}[ht]
    \includegraphics[width=\columnwidth]{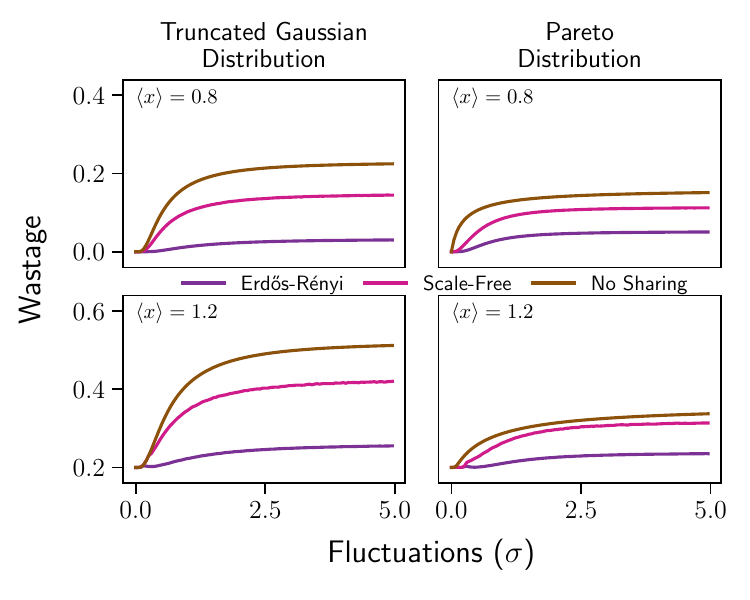}
    \caption{\label{fig:wastage_analytical} Comparison of wastage curves for ER and SF topologies for various combinations of average productions and resource generators. \emph{Left} and \emph{Right} columns correspond to the two different generators, while the \emph{Top} and \emph{Bottom} rows correspond to $\langle x\rangle < R$ and $\langle x\rangle > R$ respectively. The brown curve in each subplot shows the analytically obtained wastage when vertices are not allowed to share their surplus. As the plots show, although wastage strongly depends on $\langle x\rangle$ and the generator used, the ER network always leads to lesser wastage than the SF network.}
\end{figure}
After the surplus amounts are distributed to the neighbours, if the amount $X_i^{\text{tot}}$ on vertex $i$ is greater than its threshold $R_i$, the its fitness becomes maximum ($f_i=1$). However, out of $X_i^{\text{tot}}$, only amount $R_i$ is consumed by the vertex and the remaining amount $w_i = (X_i^{\text{tot}}-R_i)$ is wasted. It would be interesting to find out the how the total amount of wastage per unit time per vertex varies in the network. This wastage can be formally written as:
\begin{equation}
    {\small w = \lim\limits_{n\to\infty}\lim\limits_{T\to\infty}\frac{1}{nT}\sum\limits_{t=1}^{T}\sum\limits_{i=1}^n w_i(t)H(w_i(t))}
\end{equation}
where $H(x)$ denotes the Heaviside step function defined to be $1$ for $x \geq 0$ and $0$ otherwise.

Fig.~\ref{fig:wastage_analytical} compares the wastage variation with $\sigma$ for ER and SF networks with different combinations of $\langle x\rangle$ and resource generator. As these plots show, in all the cases the ER network has lower wastage compared to the SF network. This is because the degree distribution of the ER network is more homogeneous and hence the generated resource is distributed more uniformly compared to the SF network. Because of this, the ER network tends to be more fit than the SF network as we have seen in the previous section.

We can also better understand the actual wastage variation if we look at the wastage amount in the \emph{No-sharing} case. Clearly, if the vertices are not allowed to share their surplus amounts, there would be maximum possible wastage in the network. This provides us with an upper bound $W$ for $w$. For the truncated Gaussian distribution, this upper bound can be written as:
\begin{equation}
    W^{\text{gauss}} = \frac{1}{\psi(\mu, \sigma)}\int\limits_{R}^{\infty} (x-R)\exp\left(-\frac{(x-\mu)^2}{2\sigma^2}\right)dx
\end{equation}

Although this integral cannot be computed analytically, as described in the previous section, we can vary $\sigma$ and simultaneously adjust $\mu$ so that $\langle x\rangle$ remains fixed at a chosen value, and look at how the wastage $W^{gauss}$ changes by numerically computing the integral. 

We can obtain the upper bound for the Pareto distribution also, but the situation is somewhat more complicated. Here also we would like to see how the wastage varies as $\sigma$ is varied while keeping $\langle x\rangle$ at a fixed value. But as we have seen in Sec.\ref{sec:choice_prob_dists}, both $\alpha$ and $x_m$ decrease as $\sigma$ is increased. There we also saw that as long as $\sigma < \sigma_c$, $x_m > R$ which in turn means that $x > R$, and hence the expected wastage on a single vertex, when sharing is not allowed, is:
\begin{equation}
    \label{eq:wastage_pareto_less}
W^{pareto}_{(\sigma \leq \sigma_c)} = \bigintssss\limits_{x_m}^{\infty}(x-R)p(x)dx = \alpha x_m^{\alpha}\bigintssss\limits_{x_m}^{\infty} \frac{x-R}{x^{\alpha+1}}dx
\end{equation}

For $\sigma > \sigma_c$, $x_m < R$ and hence the expected wastage in absence of sharing is:
\begin{equation}
    \label{eq:wastage_pareto_more}
W^{pareto}_{(\sigma > \sigma_c)} = \bigintssss\limits_{R}^{\infty}(x-R)p(x)dx = \alpha x_m^{\alpha}\bigintssss\limits_{R}^{\infty} \frac{x-R}{x^{\alpha+1}}dx
\end{equation}

Evaluating the integrals in (\ref{eq:wastage_pareto_less}) and (\ref{eq:wastage_pareto_more}), we get the expected wastage on a single vertex as:
\begin{equation}
    W^{pareto} = \begin{cases}
\langle x\rangle - R \quad\quad\quad\quad\quad\quad \text{if}\ \sigma \leq \sigma_c\\
\\
\frac{R^{1-\alpha}}{\alpha-1}\langle x\rangle^{\alpha}\left(1-\frac{1}{\alpha}\right)^{\alpha} \quad \text{otherwise}
\end{cases}
\end{equation}
where $\alpha$ is given by Eq(\ref{eq:pareto_alpha}).

The brown curves in Fig.~\ref{fig:wastage_analytical} show the curves for the upper bound $W$. Similar to the case of network fitness, we see that the qualitative variation of $w$ for ER and SF networks is similar to that of $W$. The only effect of sharing is to scale the $W$ curve down. 

\section{Conclusion \label{conclusion}}
In this work we studied how fluctuations in the production of a resource on vertices of a network affects its fitness. Production of resource in our work is modelled by truncated Gaussian and Pareto distributions. We chose these particular distributions because they are prototypes of peaked and heavy-tailed distributions respectively. While changing the fluctuation size in both these cases we made sure that the average production on a vertex remains fixed, and hence our results are solely driven by fluctuations alone. 

We simulated our \emph{surplus distribution model} on ER and SF networks with the same average degree. We found that, independent of the average production, fluctuations in the resource production deteriorate the network fitness even when the average resource production does not change. However, we find that this worsening is limited and in the limit of infinite sized fluctuations, fitness does not go to zero but saturates to a nonzero value. We also verified a similar result of earlier works that resource dependency networks with the ER topology perform better than those with the SF topology. We have analytically obtained the lower bound on the network fitness in all the cases, and explained the sharp transitions observed for Pareto generator. Using this, we also showed how the network topology only scales up the fitness curves. We also showed how the network fitness is linked with the amount of resource wasted in the network. Furthermore we analytically computed the upper bound for the wastage in these cases. 

The main insight provided by our work is that the average resource production does not uniquely determine how good a resource dependency network performs, and that fluctuations play an important role in determining it. This is true whether the average production on each vertex is less than or greater than the threshold resource amount required by vertices. This work can be extended further in a number of directions. In this work, the production capacity of every vertex is assumed to be same. It would be interesting to see how fluctuations affect the network fitness when that is not the case. For example, production capacity can be made a function of vertex degree or of some other centrality measure like the eigenvector or betweenness. Similar to heterogeneity in the average production, the case of different fluctuation sizes for different vertices can also be explored. Moreover, in this work we focused only on the degree distribution of a network, but it is possible to study the effect of other structural properties like degree correlations, clustering, and community structure which can potentially affect the network fitness. We plan to explore these directions in future work.

\begin{acknowledgments}
SMS acknowledges funding from the DST-INSPIRE Faculty Fellowship (DST/INSPIRE/04/2018/002664) by DST India. SK acknowledges the fellowship received under the same scheme to work on the project. SMS would like to thank Sitabhra Sinha whose comment about fluctuations led SMS to conceive this project. 
\end{acknowledgments}


\end{document}